\documentclass[sigconf]{acmart}
\usepackage{amsmath}
\usepackage{tabularx}
\usepackage{array}  % for m{}
\usepackage{booktabs}  % optional for better lines
\usepackage{amsmath}   % for math environments and symbols
\usepackage{tabularx}  % 导入tabularx包
\usepackage{graphicx}  % 提供更多图形功能
\usepackage{array}  % 提供更多表格功能
\usepackage{multirow} % For multi-row and multi-column support
\usepackage{xspace}
\usepackage{booktabs}
\usepackage{tcolorbox}
\usepackage{listings}%
\usepackage{subfig}
\usepackage{float}
\usepackage{xcolor}

\newtheorem{mydef}{Definition}
\newtheorem{myProb}{\textbf{Problem}}
\AtBeginDocument{%
  }

\copyrightyear{2026}
\acmYear{2026}
\setcopyright{cc}
\setcctype{by}
\acmConference[KDD '26]{Proceedings of the 32nd ACM SIGKDD Conference on Knowledge Discovery and Data Mining V.1}{August 09--13, 2026}{Jeju Island, Republic of Korea}
\acmBooktitle{Proceedings of the 32nd ACM SIGKDD Conference on Knowledge Discovery and Data Mining V.1 (KDD '26), August 09--13, 2026, Jeju Island, Republic of Korea}
\acmPrice{}
\acmDOI{10.1145/3770854.3780334}
\acmISBN{979-8-4007-2258-5/2026/08}

\acmSubmissionID{1249}

\begin{document}

\title{Think2Go: Generative Next POI Recommendation with LLM Reasoning}

\author{Zhuang Zhuang}
\orcid{0000-0002-1227-117X}
\affiliation{%
  \institution{Dalian University of Technology}
  \department{School of Computer Science and Technology}
  \city{Dalian}
  \country{China}
}
\email{zhuang97@mail.dlut.edu.cn}

\author{Shanshan Feng}
\orcid{0000-0002-6161-9232}
\affiliation{%
  \institution{Wuhan University}
  \department{School of Computer Science}
  \city{Wuhan}
  \country{China}
}
\email{victor_fengss@whu.edu.cn}

\author{Hangwei Qian}
\orcid{0000-0003-4831-0748}
\authornote{Corresponding author.}
\affiliation{%
  \institution{A*STAR}
  \department{CFAR and IHPC}
  \country{Singapore}
}
\email{Qian_Hangwei@a-star.edu.sg}

\author{Mingqi Yang}
\orcid{0000-0002-9040-4468}
\affiliation{%
  \institution{South China University of Technology}
\department{Institute of Digital Business \& Intelligent Logistics}
  \city{Guangzhou}
  \country{China}
}
\email{yangmq@scut.edu.cn}

\author{Heng Qi}
\orcid{0000-0002-8770-3934}
\authornotemark[1]
\affiliation{%
  \institution{Dalian University of Technology}
  \department{School of Computer Science and Technology}
  \city{Dalian}
  \country{China}
}
\email{hengqi@dlut.edu.cn}

\author{Yanming Shen}
\orcid{0000-0003-4108-0230}
\affiliation{%
  \institution{Dalian University of Technology}
  \department{School of Computer Science and Technology}
  \city{Dalian}
  \country{China}
}
\email{shen@dlut.edu.cn}

\author{Baocai Yin}
\orcid{0000-0002-4164-6647}
\affiliation{%
  \institution{Dalian University of Technology}
  \department{School of Computer Science and Technology}
  \city{Dalian}
  \country{China}
}
\email{ybc@dlut.edu.cn}

\renewcommand{\shortauthors}{Zhuang Zhuang et al.}

\begin{abstract}
Next Point-of-Interest (POI) recommendation task focuses on mining user behavioral preference patterns from historical check-ins to provide personalized suggestions for the next destination. Existing methods primarily rely on shallow contextual information and handcrafted feature interactions to predict the next POI. However, the inherent sparsity and complexity of user mobility patterns limit the computational capacity of non-reasoning models to capture deep intent, while large language models (LLMs) perform suboptimally because they lack a deep understanding of semantic IDs (SIDs) when SIDs are trained separately.
%while large language models (LLMs) exhibit limited understanding of the semantic meaning of semantic IDs (SIDs) when SIDs are trained separately, leading to suboptimal performance.
To address these limitations, we propose \textbf{Think2Go}, a novel generative next POI recommendation framework, which enhances the model’s comprehension of SID representations and explores diverse spatial-temporal patterns via test-time computational scaling.
%develops a reinforcement learning framework guided by both data and model uncertainty to stimulate the reasoning capability. 
We unify supervised fine-tuning (SFT) and reinforcement learning (RL)-based reasoning within a single architecture, enabling joint optimization of memorization and adaptive reasoning to better retain user behavior patterns while exploring diverse user preferences.
 To further calibrate policy optimization in adaptive reasoning, we propose two advantage weighting mechanisms that integrate (1) \textbf{prompt epistemic uncertainty}, estimated via kernel density methods to assess the spatial-temporal periodic pattern alignment between queries and user history, promoting increased exploration under high epistemic uncertainty; and (2) \textbf{reward-informed advantage scaling}, captured by normalizing rewards against their maxima to adapt update magnitudes, thereby improving training stability and mitigating overfitting to noisy signals. This joint calibration forms an implicit curriculum learning strategy, delivering fine-grained, instance-aware policy updates that prevent entropy collapse and support robust exploration. Extensive experiments conducted on three real-world datasets demonstrate that Think2Go exhibits strong generalization capabilities and enhances the LLM’s understanding of SIDs.

%Specifically, we initialize the spatial-temporal data confusion of the target location prediction based on the periodicity and geographical distribution of historical trajectories. Meanwhile, the model’s prediction confidence over sampled data is measured by the gap between its output and the maximum reward value. They both work together to maintain high policy entropy, enabling continuous exploration and exploitation of diverse strategy evolution.
% Under this design, our model achieves state-of-the-art performance on three public datasets.

%which dynamically adjusts the extent of policy updates. 
%which explicitly leverages the uncertainty in spatial-temporal mobility patterns in the data and the correctness in the model to modulate the magnitude of policy updates. 

\end{abstract}

\begin{CCSXML}
<ccs2012>
   <concept>
       <concept_id>10002951.10003227.10003236.10003101</concept_id>
       <concept_desc>Information systems~Location based services</concept_desc>
       <concept_significance>500</concept_significance>
       </concept>
   <concept>
       <concept_id>10002951.10003227.10003351</concept_id>
       <concept_desc>Information systems~Data mining</concept_desc>
       <concept_significance>500</concept_significance>
       </concept>
   <concept>
 </ccs2012>
\end{CCSXML}

\ccsdesc[500]{Information systems~Location based services}
\ccsdesc[500]{Information systems~Data mining}
\ccsdesc[500]{Human-centered computing~Ubiquitous and mobile computing design and evaluation methods}
\keywords{POI Recommendation, Generative Model, Reasoning Model, LLM}


\maketitle
\newcommand\kddavailabilityurl{https://doi.org/10.5281/zenodo.18206701}
\ifdefempty{\kddavailabilityurl}{}{
\begingroup\small\noindent\raggedright\textbf{Resource Availability:}\\
% please change the following context to include multiple artifacts if necessary, including data, models, code, etc.
The source code of this paper has been made publicly available at \url{\kddavailabilityurl}.
\endgroup
}

\section{Introduction}
Location-based social networks (LBSNs) are increasingly becoming an integral part of modern life, as a growing number of users share their check-in records at interesting points of interest (POIs) via mobile platforms such as Foursquare and Google Maps. As a result, POI recommendation has emerged as a key research task~\cite{ruan2024twitter, wan2025poi} based on LBSN data, aiming to understand users’ behavioral preferences~\cite{wongso2024genup, zhuang2024tau} from their historical trajectories and provide location-based services~\cite{zhuang2021traffic, li2024geo, li2025pcde}. A notable variant of this task is next POI recommendation, which aims to predict the next location a user is likely to visit based on the spatial-temporal context of their check-in history and a given temporal cue~\cite{yu2025bigcity}, assisting users in trip planning~\cite{xie2024travelplanner,li2024more} and in discovering locations of potential interest~\cite{yang2020location, tang2024itinera, zhuang2024cmaan}.

Recently, large language models (LLMs) have demonstrated remarkable progress in contextual understanding~\cite{hao2025oxygenrec}, marking a significant breakthrough in tackling complex real-world tasks. This has motivated the use of LLM-based generative modeling as a novel paradigm for next POI recommendation, where the model directly generates the ID of the next POI at the target time based on user-provided prompts. Compared to traditional POI recommendation methods, this approach avoids the semantic loss that occurs when heterogeneous textual data are converted into numerical IDs. It effectively leverages the general knowledge of LLMs to capture contextual dependencies in user check-in trajectories. Moreover, semantic POI ID representations have been proposed to initialize IDs with similar meanings for geographically or functionally related locations, further enhancing the LLM's ability to understand and generalize POI semantics. 

Recent methods~\cite{li2024large,long2024unleashing,li2024limp, liu2024nextlocllm} based on LLMs primarily focus on fine-tuning pre-trained models but overlook the benefits of expanding computational capacity through test-time scaling, enhancing the next POI generation via reasoning. For instance, GNPR \cite{wang2025generative} fine-tunes the base model with pre-trained SIDs to enforce semantic alignment between the model and the ID space. However, this decoupled training strategy leads to an incomplete understanding of SID information, particularly under sparse POI recommendation scenarios. As a result, current LLM-based approaches struggle to accurately incorporate SID signals during generation, yielding suboptimal performance. This insufficient comprehension of semantic IDs is further reflected in the model’s poor generalization on out-of-domain datasets and its limited capability to distinguish SID categories.

These limitations highlight the importance of designing effective reasoning models to improve both SID understanding and sequential modeling within LLMs. However, developing a generation-based POI recommendation framework with integrated reasoning is nontrivial due to several challenges:
(1) \textbf{Model Architecture:}
Traditional reasoning models rely on RL-based policy optimization, which assumes that the base model has sufficient predictive capability to sample correct answers. Moreover, overfitting during the supervised fine-tuning (SFT) stage can restrict exploration in subsequent RL training and decouple memorization from reasoning, ultimately resulting in suboptimal learning \cite{liu2025uft,huang2025blending,wei2025oneloc}.
%Conventional training pipelines—where supervised fine-tuning (SFT) is followed by RL reasoning—can hinder reasoning performance. Overfitting during SFT may constrain exploration and decouple memorization from reasoning, ultimately leading to suboptimal learning.
(2) \textbf{Prompt-Aware Adaptive Optimization:}
Existing policy optimization methods treat all prompts equally, neglecting varying prediction difficulties, which limits exploration of more challenging prompts. In POI recommendation scenarios, predicting targets that deviate from a user’s habitual temporal patterns or commonly visited regions presents greater difficulty.
(3) \textbf{Model-Aware Adaptive Optimization:}
While Dr.GRPO \cite{liu2025understanding} removes the variance term in advantage estimation to alleviate signal amplification caused by estimation bias, it still overlooks another form of bias: the tendency of the model to produce similar intra-group advantage distributions across samples of varying difficulty. 
For example, a hard instance with rewards [2.5, 2.5, 1.5] and an easy one with [4.5, 4.5, 3.5] both yield the same advantage [0.33, 0.33, –0.67] under mean-centered advantage estimation. This leads to uniform updates across tasks of different difficulty, which undermines fine-grained learning and limits the ability to assign appropriate update magnitudes based on model competence.

\begin{figure}
    \centering
    \includegraphics[width=0.90\linewidth]{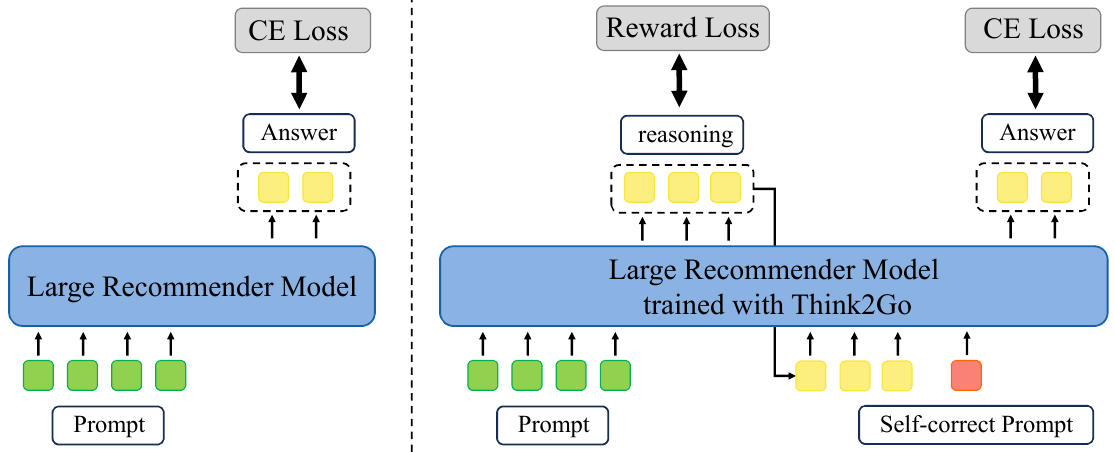}
    \caption{The architecture overview of Think2Go, which unifies RL-based reasoning and SFT within the auto-regressive generation process, connecting reasoning and supervised answers through self-correction.}
    \label{fig:intro}
\end{figure}

To address the aforementioned challenges, we propose Think2Go as shown in Figure~\ref{fig:intro}, a unified framework that combines SFT and RL-based reasoning to unlock LLM’s test-time reasoning capacity, thereby improving its comprehension of semantic identifiers and modeling of sequential user behaviors.
To tackle the first challenge, we introduce a unified training paradigm that integrates both SFT and RL-based policy optimization. The model generates structured outputs in the form of \textit{Think->Go->Self-Correct->Answer}, where the \textit{Self-Correct} token serves as a bridge between the reasoning and SFT phase. The final \textit{Answer} token is used for computing the SFT loss, while the \textit{Think} and \textit{Go} segments comprise the reasoning trajectory, which is optimized via policy gradients. This unified structure facilitates seamless learning between memorization and exploration.
To address the second challenge, we propose an epistemic uncertainty advantage calibration strategy for prompt-based learning. Specifically, we use kernel density estimation to assess the spatial-temporal similarity between the current query and the user’s historical behavior. This similarity quantifies epistemic uncertainty and adaptively modulates the advantage signal to better explore challenging instances.
For the third challenge, we introduce difficulty-aware advantage weighting, which regularizes the reward of each sample by its maximum value to reflect prediction difficulty. This allows the model to adaptively adjust the scale of policy updates. As a result, the combined effect of prompt and model-aware calibrations serves as an implicit curriculum learning mechanism, delivering fine-grained, instance-specific gradients that enable more effective and adaptive policy optimization.

Our contributions can be summarized as follows:
\begin{itemize}
\item We propose Think2Go, the first generative next POI recommendation framework based on LLM reasoning, which unifies SFT and RL-based reasoning to better balance memorization and exploration.

\item We propose an advantage calibration mechanism that adapts to both prompt epistemic uncertainty and prediction difficulty, implicitly forming a competence-aware curriculum that refines exploration and enhances learning efficiency.  

\item Think2Go consistently outperforms baselines across three public datasets. Further evaluation on out-of-domain datasets and SID classification tasks shows that reasoning improves LLMs' understanding of SIDs and enhances generalization. %Our implementation is available at \url{https://github.com/zhuang999/Think2Go_KDD26}.

\end{itemize}

\section{Related Work}
\label{related_work}
\subsection{Next POI Recommendation}
Existing literature on next POI recommendation primarily focuses on modeling users' historical check-in sequences to capture the spatial-temporal transition patterns~\cite{wang2023eedn, sun2025exploring, qiao2025hyperbolic} between POIs, making it a typical sequential recommendation task~\cite{jiang2024trimlp,rao2025disentangled, zhuang2025mgstdn,jiang2025auto,zhang2025onetrans}. Early studies~\cite{cheng2013you,feng2015personalized,mathew2012predicting} widely adopted Markov Chain (MC) methods~\cite{chung1967markov}, leveraging personalized transition matrices to model user preferences and mobility patterns. However, these approaches fall short in capturing users' complex interests and implicit behavioral signals compared to deep learning-based models.

With the rise of deep learning, research has gradually shifted towards models based on Recurrent Neural Networks (RNNs)~\cite{hochreiter1997long} and attention mechanisms~\cite{vaswani2017attention}. For instance, DeepMove~\cite{feng2018deepmove} integrates attention and gating mechanisms to jointly learn users’ long-term periodic and short-term preferences. STGN~\cite{zhao2020go} enhances LSTM architectures by incorporating spatial-temporal gates into the gating units, enabling the modeling of long- and short-term spatial-temporal preferences. STAN~\cite{luo2021stan} and STiSAN~\cite{wang2022spatial} further introduce attention mechanisms to capture both continuous and non-continuous spatial-temporal contextual intervals in the modeling process.
In addition, Graph Neural Networks (GNNs) have been leveraged to model high-level POI transition patterns by utilizing the user-POI graph structure~\cite{qiao2023rotdiff,lyu2024cross}. For example, Graph-Flashback~\cite{rao2022graph} constructs a powerful weighted user-POI knowledge graph to initialize POI embeddings, while STHGCN~\cite{yan2023spatio} employs hypergraphs to capture trajectory-level collaborative signals.

Recently, LLMs with extensive commonsense knowledge have shown strong potential in recommendation tasks, leading to a rise in LLM-based approaches for next POI recommendation~\cite{liu2025geography}. LLM-Mob~\cite{wang2023would} enhances the modeling of historical trajectory context through in-context learning. NextLocLLM~\cite{liu2024nextlocllm} attempts to predict spatial coordinates directly to improve spatial reasoning within LLMs. LLM4POI~\cite{li2024large} reformulates next POI recommendation as a question-answering task and fine-tunes the model using QLoRA-based supervised learning. Refine-POI~\cite{li2025refine} applies Group-based Rule-guided Policy Optimization (GRPO)~\cite{shao2024deepseekmath} with recommendation-driven reward design to generate top-k recommendation lists. However, they have not explored reasoning-enhanced models to boost generalization. While Refine-POI leverages the GRPO framework for reasoning-based prediction, it overlooks the optimization of the advantage function, which limits performance improvements.
  
\subsection{Reinforcement Learning for LLM Reasoning}
Recent advancements in generative Artificial Intelligence (AI), particularly with models like OpenAI-o1~\cite{openai2024reasoning} and Deepseek-R1~\cite{jin2025search}, have extended the scope of research from large language models (LLMs) to large reasoning models (LRMs). This shift is predicated on the application of the Reinforcement Learning from Human Feedback (RLHF) training framework, coupled with test-time scaling and an emphasis on deep thinking capabilities. The core innovation lies in explicitly leveraging a long chain-of-thought (CoT) mechanism~\cite{liu2025onerec}. This encourages models to generate a substantial number of intermediate reasoning tokens before arriving at a final answer, leading to significant performance improvements in complex domains such as mathematics, code generation, and question answering.

GRPO has emerged as a leading reinforcement learning algorithm for training reasoning-augmented large language models GRPO eliminates the need for a separate value model by employing group-level rewards based on rule validation, thereby enabling self-play validation to enhance reasoning efficacy~\cite{deng2025onerec}. However, GRPO faces challenges such as inefficient sampling and length bias, which have spurred further optimizations in algorithms like Decoupled Advantage Policy Optimization (DAPO)~\cite{yu2025dapo} and Dr.GRPO~\cite{liu2025understanding}. Large Reasoning Models have garnered considerable interest in the domain of generative recommendation systems. R$^2$EC \cite{you2025text} re-conceptualizes recommendation model architectures by integrating an inherent reasoning capability within a regression process. R$^2$Rec~\cite{zhao2025reason} employs sampled interaction chains from user-item graphs to guide chain-of-thought reasoning. Nevertheless, both overlook sparse spatiotemporal behavior patterns, limiting their policy optimization and overall performance.

\section{Preliminaries}
\label{sec: preliminaries}

\subsection{Key Concept and Problem Definitions\label{sec:notations}}
\begin{mydef}[Check-in and Trajectory Sequence]
Each check-in record is defined as a triplet $s_i=(u, l_i, t_i)$, where user $u$ visits a POI $l_i$ at timestamp $t_i$. Each POI $l_i$ is characterized by its geographic coordinate $g_i = (latitude = {lat_i}, longitude = {lon_i})$, along with an associated categorical label $c_i$ indicating its type. Here, we denote the set of all users as $U=\{u_1,u_2,\ldots,u_{|U|}\}$, and the set of all POIs as $L=\{l_1,l_2,\ldots,l_{|L|}\}$.
\end{mydef}

\begin{myProb}[Next POI recommendation]
Given a user $u_i$, the trajectory $S_u = \{ s_1,s_2,\ldots,s_n\}$ denotes the sequence of historical check-ins sequence from time step $t_1$ to $t_n$. The objective of the next POI recommendation task is to identify the next POI $l_{n+1}  \in L$ that the user $u$ is most likely to visit in the next time step $t_{n+1}$.
%and to determine the desired output $l_{n+1}  \in L$ for the user's subsequent check-in.
\end{myProb}

\subsection{Kernel Density Estimates}
\label{sec:kde}
Given the spatial $G_u=\{g_i\}_{i=1}^n$ and temporal $T_u=\{t_i\}_{i=1}^n$ features extracted from a user's historical check-in sequence $S_u$, a query $q^G$ or $q^T$ representing the spatial or temporal attributes of a target POI, we employ a kernel density estimation (KDE) function defined as $k(x, x') = \alpha \exp\left( -\gamma (x - x')^2 \right)$, where $\alpha$ determines the amplitude of the kernel function, and $\gamma$ controls the smoothness of the similarity measure. This formulation quantifies the spatial-temporal uncertainty of a query by estimating its likelihood under the user’s historical distribution. A lower density score indicates that the query falls in a sparse region of past behavior, suggesting higher uncertainty or more exploratory intent.

\subsection{Reinforcement Learning for Large Reasoning Models}
Inspired by DeepSeek-R1, it is worth noting that RL-based post-training methods enhance reasoning capabilities by promoting discriminative exploration from high-quality reasoning trajectories. 
In particular, GRPO introduces a novel approach that samples $G$ groups of rollouts ${o^1, o^2,\dots,o^G}$ from the old policy $\pi_{\theta_{\text{old}}}$ for each input $q$ and utilizes normalized rule-based rewards to estimate advantages $\hat{A}$, thereby guiding the current policy $\pi_{\theta}$ optimization. 
Following GRPO, DAPO further improves sample efficiency by removing the KL penalty and incorporating dynamic sampling, clip-higher updates, and token-level policy gradients. The overall objective can be formalized as follows:

\begin{equation}
\begin{split}
    \mathcal{L}_{\mathrm{DAPO}}(\theta) = & \,
    \mathbb{E}_{(q,a) \sim D, \{ o^i \}_{i=1}^G \sim \pi_{\theta_\mathrm{old}}(\cdot|q)} 
    \Bigg[ \frac{1}{\sum_{i=1}^G |o^i|} \sum_{i=1}^G \sum_{t=1}^{|o^i|} \\
    & \min \Big( r_t^{(i)}(\theta) \hat{A}^i_t, 
    \mathrm{clip}\big( r_t^{(i)}(\theta), 1 - \epsilon_\mathrm{low}, 1 + \epsilon_\mathrm{high} \big) \hat{A}^i_t \Big) \Bigg], \\
    \text{s.t.} \quad & 0 < \left| \{ o^i \mid \mathrm{is\_equivalent}(a, o^i) \} \right| < G
\end{split}
\label{eq:dapo}
\end{equation}
where $D$ denotes a dataset of queries $q$ paired with corresponding ground-truth answers $a$. The clipping parameters $\epsilon_\mathrm{low}$ and $\epsilon_\mathrm{high}$ define the trust region for importance ratio clipping, which contributes to optimization stability. The importance ratio at time step $t$ for trajectory $i$ is defined as $r_t^{(i)}(\theta) = \frac{\pi_{\theta}(o_t \mid q, o_{<t})}{\pi_{\theta_{\text{old}}}(o_t \mid q, o_{<t})}$, measuring the relative likelihood of the token under the updated policy $\pi_\theta$ versus the old policy $\pi_{\theta_{\text{old}}}$. And, the estimated advantage $\hat{A}_i$ is calculated as follows:
\begin{equation}
\hat{A}_i = \frac{R_i - \mathrm{mean}(\{R_i\}_{i=1}^{G})}{\mathrm{std}(\{R_i\}_{i=1}^{G})}, 
\end{equation}

%\quad \text{where} \quad R_i = 
% \begin{cases}
% 1.0 & \text{if } \texttt{is\_equivalent}(a, o^i) \\
% 0.0 & \text{otherwise}
% \end{cases}

\begin{figure*}
    \centering
    \includegraphics[width=0.90\linewidth]{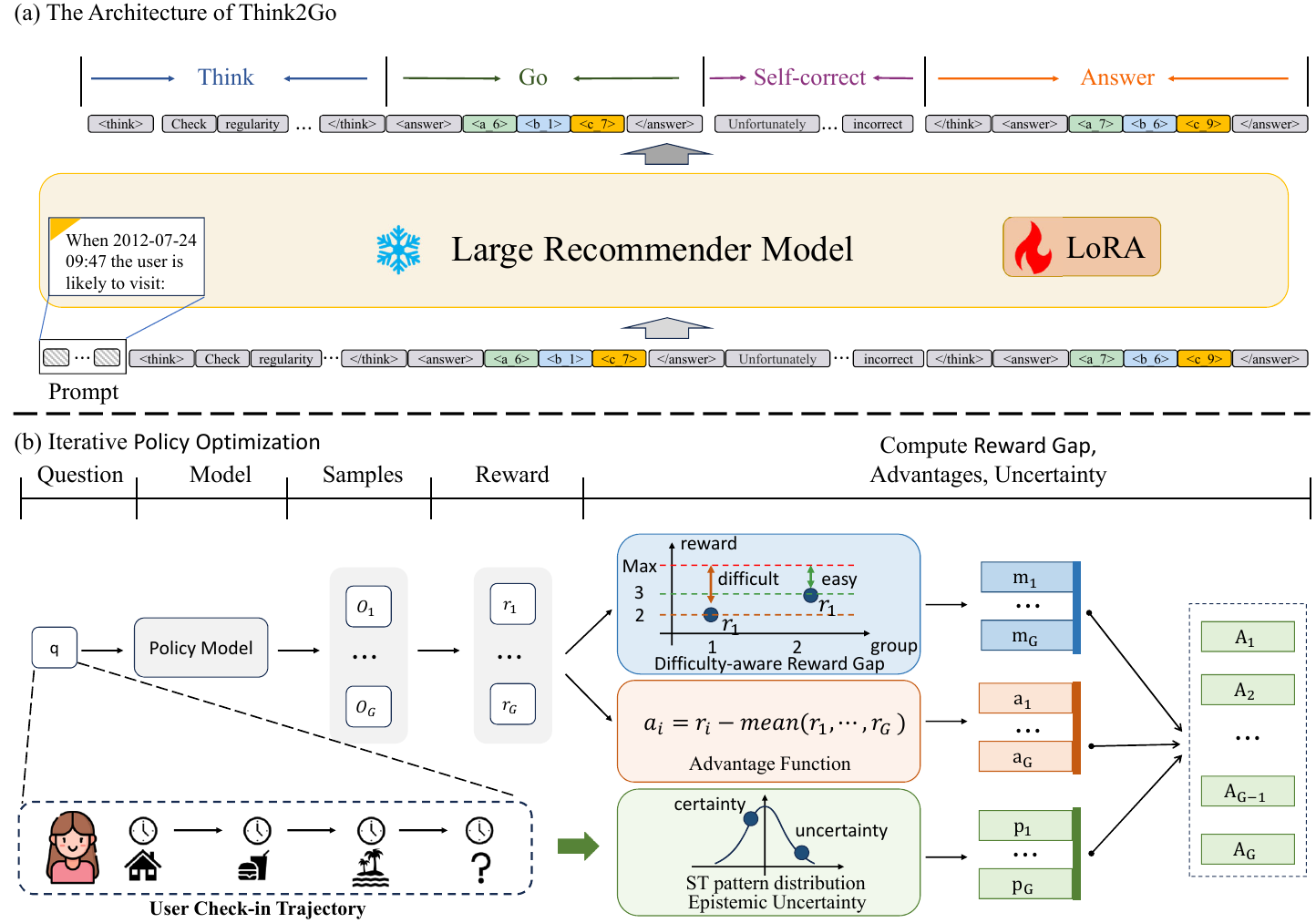}
    \caption{The overall framework of Think2Go, which performs iterative policy optimization over the \textit{Think→Go} token sequence, while the \textit{Answer} tokens are trained via SFT. To enhance iterative policy learning, we propose a Difficulty-Aware Advantage Weighting Estimator to reflect reasoning difficulty via reward gaps, and a Spatial-Temporal Epistemic Uncertainty (STEU) Estimator based on kernel density estimation to capture epistemic uncertainty. These jointly refine a centralized advantage function, enabling more precise policy gradient updates.}
    \label{fig:architecture}
\end{figure*}

\section{Methods}
\label{sec:methods}
Think2Go is a large next POI recommendation model that adopts a “think-then-recommend” paradigm. Section~\ref{sec:architecture} presents the architecture, wherein generative reasoning and recommendation are jointly integrated into a unified framework. Section~\ref{sec:optimization} introduces a tailored RL algorithm designed to alleviate entropy collapse and maintain an effective balance between exploration and exploitation.

\subsection{Model Architecture}
\label{sec:architecture}
Our proposed Think2Go model is built upon a decoder-only backbone, leveraging self-correction prompts as a bridge to unify supervised fine-tuning (SFT) and reinforcement learning (RL). 
In existing RLHF approaches, they typically perform SFT followed by RL, which limits learning efficiency and leads to catastrophic forgetting from SFT\cite{liu2025uft,huang2025blending}. To overcome this, we unify SFT and RL within a single architecture. 
It employs SFT to ensure the model retains correct answers, while RL is used to enhance the model’s ability to explore and infer correct answers through reasoning. 
As illustrated in Figure~\ref{fig:architecture}, the old policy model first generates a reasoning trace composed of a Think step and a final prediction (Go). While the correct answer is typically appended for SFT, we instead introduce a self-correct prompt to provide contextual grounding. This forms a \textbf{Think–>Go–>Self–correct–>Answer} structure, where tokens from the reasoning trace are used for RL, and tokens from the appended answer are used for SFT.

Specifically, SFT uses the standard next-token prediction task, with its objective function defined as follows:
\begin{equation}
    \mathcal{L}_{\text{SFT}} = - \sum_{(x, y) \in \mathcal{D}} \sum_{i=1}^{|y|} \log P_\theta (y_i \mid x, y_{<i}),
\end{equation}
where $y_i$ denotes the $i$-th token in the answer $y$, and $P_\theta (y_i \mid x, y_{<i})$ denotes the probability of predicting $y_i$ conditioned on the previously generated tokens. The final model optimization loss is formulated as $\mathcal{L} = \mathcal{L}_{\text{SFT}} + \alpha\mathcal{L}_{\text{CAPO}}$, where $\alpha$ is a hyperparameter balancing the SFT loss and the policy optimization loss. And, $L_{CAPO}$ will be detailed in Section~\ref{sec:optimization}.

\subsection{Calibrated Advantage Policy Optimization (CAPO)}
\label{sec:optimization}
\subsubsection{Adaptive Advantage Calibration in Robust Policy Optimization}
Adaptive advantage calibration facilitates self-adjusting policy updates in Think2Go. Specifically, we customize a sample-level calibration coefficient $c^i_t$ and a token-level calibration coefficient $e^i_t$. 
This mechanism leverages both epistemic uncertainty and the model’s predictive capability to realize an implicit curriculum learning effect. It dynamically balances exploration and exploitation: challenging samples with high epistemic uncertainty and correct predictions receive larger updates to encourage exploration, while incorrect predictions are updated conservatively to ensure robustness. As training progresses, the model naturally allocates more learning capacity to correctly predicted samples and epistemically challenging problems, thereby accelerating convergence and enhancing overall generalization.

To better understand this mechanism, we present a simplified gradient analysis, considering the unclipped loss function for clarity:
\begin{equation}
L_i(\theta) = r^{(i)}_t(\theta) \cdot \hat{A}_i = \frac{\pi_\theta(o_i \mid q)}{\pi_{\theta_{\text{old}}}(o_i \mid q)} \cdot \hat{A}_i,
\end{equation}
Accordingly, the policy update with the global learning rate $\eta$ is given by:
\begin{equation}
\theta \leftarrow \theta + \eta \cdot \nabla_{\theta} \log \pi_{\theta}(o_i \mid q) \cdot r^{(i)}_t(\theta) \cdot \underbrace{\left[c^i_t \cdot (A_i + e^i_t)\right]}_{\hat{A}_{i}},
\end{equation}
where $e^i_t$ represents the entropy of each reasoning token, which is used to measure its uncertainty, and is defined as follows:
\begin{equation}
e^i_t = - \sum_{j=1}^{|V|} p^i_{t,j} \log p^i_{t,j}, \quad \text{where} \quad p^i_t = \pi_\theta(\cdot \mid q^i, o^i_{<t})
\end{equation}
where $|V|$ denotes the vocabulary size of the policy model. They show that the value of the advantage directly influences the magnitude of policy updates. Among them, the incorporation of token-level entropy encourages exploration and promotes policy diversity during reasoning, particularly for uncertain tokens \cite{wang2025beyond}.
Moreover, we integrate calibration coefficients into the advantage estimation to realize prompt-specific adaptive learning rates, enabling dynamic adjustment of update magnitudes based on both the epistemic uncertainty of the prompt and the model’s predictive competence.

%we dynamically scale the gradient updates according to the cognitive difficulty of the input and the model’s prediction capability. This design enables a question-specific adaptive learning rate through the joint calibration mechanism, dynamically modulating the update magnitude based on the epistemic uncertainty of the data and the model’s predictive competence. 

\subsubsection{Rule-based Reward Estimation}
Given the sampled trajectories, the next step is to assign appropriate rewards to each one. Conventionally, reward signals include format rewards and evaluation-based rewards (e.g., Acc@1), which encourage the model to generate outputs with correct answer formatting and ultimately improve task performance.
However, the inherent sparsity of such rewards severely restricts effective optimization. This is particularly problematic in the initial training phase, where the model often fails to generate outputs in the prescribed format and much less the correct answer.

To address this challenge, we design a progressive reward mechanism that first encourages the model to generate well-formatted responses and then to produce semantically correct answers.
For format rewards, we use a two-level scheme: if the output contains a partial format (i.e., \texttt{<answer>:}), a reward of 0.5 is assigned; if the output satisfies the full structure \texttt{</think><answer>:SID</answer>} and exceeds a length threshold of 300 tokens, a full reward of 1.0 is given. Benefiting from SFT, the model gradually learns to produce partially correct formats before mastering the complete structure. The length constraint is crucial for preventing reward hacking, where the model exploits formatting alone to obtain undeserved rewards.
For evaluation-based rewards, we adopt a fine-grained scoring strategy based on decomposed semantic ID units. Each correctly predicted unit receives a reward of $\frac{3}{L}$, where $L$ denotes the total number of semantic units. This approach provides graded feedback for partially correct answers and encourages the model to incrementally refine its output toward complete correctness.

\subsubsection{Spatial-Temporal Epistemic Uncertainty (STEU)}
Group-wise normalized advantage estimation captures the model’s relative competence on given prompts and serves as the basis for subsequent policy updates. However, this uniform treatment fails to account for the varying difficulty of different instances, resulting in equally weighted updates regardless of prediction complexity. Such insensitivity may diminish learning incentives for challenging samples, thereby constraining the model’s exploration capacity and generalization ability. For the next POI recommendation, where prediction tasks are inherently heterogeneous due to spatial preferences and temporal regularities in user behavior, ignoring epistemic uncertainty distinctions can prevent the model from fully optimizing on high-value, hard-to-learn samples, ultimately limiting overall performance.

To address the aforementioned issue, we employ a kernel density estimator in section~\ref{sec:kde} to quantify the spatial-temporal uncertainty of each trajectory, which serves as an indicator of the question’s epistemic difficulty. 
Firstly, from the temporal perspective, we consider the periodic characteristics of user behavior by measuring the temporal differences between the target query time and the user’s historical check-in times. A Gaussian kernel is used to perform smooth density estimation over these time differences, capturing the confidence of the query time under the user’s historical temporal distribution. To ensure normalization and interpretability as confidence, we take the average of all kernel responses. The formulation is:
\begin{equation}
p_{\text{temporal}}(q) = \frac{1}{N} \sum_{i=1}^{N} \exp\left(-\frac{\left[\min\left(|t_q - t_i|,\ 24 - |t_q - t_i|\right)\right]^2}{2\sigma^2}\right),
\end{equation}
where $t_q$ is the target query timestamp, $t_i$ denotes the timestamp of the $i$-th historical check-in, $\sigma$ is the bandwidth of the Gaussian kernel, and $N$ is the number of historical check-ins. %A higher value of $\text{p}_{\text{time}}(q)$ indicates that the query time is more consistent with the user's past behavior, implying lower temporal uncertainty.

Second, from the spatial perspective, we quantify whether the target query location lies within the user's frequently visited regions by measuring the geodesic distance between the query location and historical check-in points. A Gaussian kernel is applied to these distances to estimate a spatial preference score, reflecting the user's activity density around the query location. The final spatial confidence is obtained by averaging these kernel values:
\begin{equation}
p_{\text{spatial}}(q) = \frac{1}{N} \sum_{i=1}^N \exp\left(-\frac{\Delta D(l_q, l_i)^2}{2\sigma^2}\right),
\end{equation}

where \(l_q\) is the query location, \(l_i\) denotes the \(i\)-th historical check-in location, \(\sigma\) is the kernel bandwidth, and \(N\) is the number of historical check-ins.
A higher $p_{\text{spatial}}(q)$ or $p_{\text{temporal}}(q)$ indicates that the query is more consistent with the user's past behavior, implying lower uncertainty. Finally, we combine them using a hyperparameter $\beta$ to form the query epistemic uncertainty $p=\beta \cdot \bigl(1- p_{\text{temporal}}(q)\bigr) + (1-\beta) \bigl(1- p_{\text{spatial}}(q)\bigr)$.

\subsubsection{Difficulty-aware Reward Gap (DRG)}  
Dr.GRPO removes the regularization term to avoid difficulty bias caused by excessively small group-wise standard deviations when the input prompts are either too hard or too easy, and retains only the centered outcome reward as the advantage. However, this leads to a problem in multi-reward settings where predictions of varying quality receive identical advantages if their rewards follow similar group-wise distributions. As a result, it weakens the model's ability to distinguish between better and worse strategies, reduces the directional specificity of gradient updates, and increases the risk of converging to suboptimal local minima. 

For example, consider a group size of 3, where the reward values are [4.5, 4.5, 3.5] for an easy instance and [2.5, 2.5, 1.5] for a hard one. After mean-centering, both groups yield the same advantage distribution: [0.33, 0.33, -0.67]. This identical advantage assignment leads the model to perform equivalent policy updates for both groups, disregarding the absolute magnitude of the rewards and failing to capture the intrinsic difficulty of the underlying prediction task. As a result, the optimization process becomes insensitive to task difficulty, weakening the learning incentives for challenging samples and ultimately constraining the effectiveness of policy optimization and the upper bound of model performance.

To mitigate the value dilution introduced by intra-group normalization, we re-scale the group-wise advantage using the original reward magnitude relative to the maximum reward to preserve the signal of absolute performance. This adjustment mitigates the loss of reward intensity information caused by standardization alone. Our DRG is defined as:%This adjustment restores reward signal intensity across groups, ensuring that higher-value samples receive proportionally greater gradient updates. 
\begin{equation}
    m=1 - \frac{R_{max}-R}{R_{max}},
\end{equation}
where $R$ denotes the current sample reward, $R_{max}$ denotes the maximum reward value. As illustrated in the earlier example, the DRG advantage calibration coefficients are 
[1.0,1.0,0.778] for the easy instance and [0.556,0.556,0.333] for the difficult one, reflecting the intended sensitivity to different samples. Serving as scaling factors based on the centered advantage, these coefficients enable the differentiation of instance difficulty, leading to more fine-grained and adaptive policy optimization.
Through comparative experiments, we further observe that it effectively alleviates entropy collapse by maintaining a high entropy level for the reasoning tokens, which implies greater exploration capacity.

%We propose a Reward-Aware Advantage Reweighting (RAAR) mechanism to reintroduce sample-wise reward intensity into group-normalized advantage estimation, thereby alleviating value dilution and improving gradient alignment with task difficulty.

\subsubsection{Training Objective}
With the spatial-temporal epistemic uncertainty and the difficulty-aware advantage weight, we proceed to compute the policy optimization loss. However, as noted in Dr.GRPO, conventional advantage regularization introduces bias, where the variance term may disproportionately amplify the advantage signal. To mitigate this issue, we eliminate the variance component. Consequently, the final advantage in Eq.~\ref{eq:dapo} is updated as follows:

\begin{align}
\hat{A}^i_t &= e^{(\gamma \cdot p \cdot m)} \cdot A^i_t, \\
A^i_t &= R_i - \mathrm{mean}(\{R_i\}_{i=1}^{G}),
\end{align}
$\gamma$ is a hyperparameter that controls the scaling effect of the calibration coefficient on the advantage.

\section{Experimental Setting}
\label{sec: settings}
 
\subsection{Datasets} 
We evaluated our model's performance on three widely used real-world datasets. Foursquare-NYC~\cite{yang2014modeling} and Foursquare-TKY~\cite{yang2014modeling} contain check-in data from New York City and Tokyo, respectively, from April 2012 to February 2013. Gowalla-CA~\cite{cho2011friendship} collected check-ins from February 2009 to October 2010 in California and Nevada. Each check-in record consists of userID, POIID, POI Category, Longitude, Latitude, and Timestamp.
We maintained the following preprocessing techniques, consistent with GNPR:
(i) Filtered out unpopular POIs with fewer than ten interactions.
(ii) Removed check-in records of inactive users with fewer than ten check-ins.
(iii) Removed users and POIs from the test set that did not appear in the training set to ensure consistency between training and evaluation.
Then, check-in records were grouped by user and sorted chronologically. The first 80\% of these records were used for training, the subsequent 10\% for validation, and the final 10\% for testing. For each user, the last check-in was used as the answer, while the previous historical check-ins were used to build the input sequence for the problem. Table~\ref{tab: statistics} shows the check-in data statistics for the three datasets used in our experiments.

\begin{table}
  \caption{Statistics of three benchmark datasets.}
  \label{tab: statistics}
  \begin{tabular}{l>{\centering\arraybackslash}p{0.12\linewidth}>{\centering\arraybackslash}p{0.12\linewidth}>{\centering\arraybackslash}p{0.12\linewidth}}
    \toprule
    \textbf{Dataset}&\textbf{NYC}& \textbf{TKY} &\textbf{CA}\\
    \midrule
    Categories& 209& 190&301\\
    POIs& 5,135& 7,873&14,027\\
    Users& 1,083& 2,293&6,592\\
    % Check-in  &     &       & \\
 Avg. Check-In & 136& 195&53
\\
    Fre. Check-In& 68& 100& 10\\
    \bottomrule
  \end{tabular}
\end{table}

\subsection{Baselines}
We have carefully selected the following state-of-the-art methods as our baselines, including those based on LLM methods:
(1) \textbf{Traditional methods}:
\texttt{PRME}~\cite{feng2015personalized}, a metric embedding method to learn personalized user transition patterns.
\texttt{LSTM}~\cite{graves2012long}, a sequence recommendation model based on recurrent neural networks.

(2) \textbf{Sequence-based methods}:
\texttt{STAN}~\cite{luo2021stan}, a self-attention model that considers spatio-temporal intervals.
\texttt{TPG}~\cite{luo2023timestamps}, a transformer model that uses timestamps as prompts.
\texttt{ROTAN}~\cite{feng2024rotan}, a rotation-based temporal attention model that incorporates a temporal rotary positional encoder to enhance temporal awareness.

(3) \textbf{GCN-based methods}:
\texttt{GETNext}~\cite{yang2022getnext}, a graph convolutional network model, leverages a global transition graph to capture collaborative signals from other users.
\texttt{STHGCN}~\cite{yan2023spatio}, a hypergraph-based model that captures both inter-user and intra-user relationships.

(4) \textbf{LLM-based methods}:
\texttt{LLM-Mob}~\cite{wang2023would}, a reasoning-driven framework utilizing context-inclusive prompts to capture multi-scale mobility dependencies.
\texttt{LLM4POI}~\cite{li2024large}, a supervised fine-tuned large language model based on question–answer pairs.
% \texttt{$R^2$EC}~\cite{}, an autoregressive recommendation model with intrinsic reasoning capabilities.
\texttt{GNPR}~\cite{wang2025generative}, a generative recommendation LLM based on semantic IDs.

(5) \textbf{Reasoning-based methods}:
\texttt{Refine-POI}~\cite{li2025refine}, an improved reasoning model based on GRPO.

\subsection{Evaluation Metrics}
Following previous work and use the widely adopted evaluation metric, Accuracy@K ( Acc@K) and Mean Reciprocal Rank (MRR), to assess the performance of comparative experiments. Acc@K measures the hit rate within the top-$K$ results, whereas MRR evaluates the specific rank of the ground-truth POI. Specifically, the Acc@K metric is formally defined as follows:
\begin{equation}
    \text{Acc@K} = \frac{1}{|U|} \sum_{u \in U} \frac{|S^K_u \cap S_u|}{|S_u|}, 
    \label{eq:acc}
\end{equation}
where $S_u$ is the target POI, and $S^K_u$ denotes the list of top-$K$ predicted POIs. This metric measures the proportion of correctly sampled items within the top-K predicted samples. And, $MRR$ is defined as follows:
\begin{equation}
    MRR = \frac{1}{{|Test|}}\sum\limits_{i = 1}^{|Test|} { \frac{1}{rank_i}}
\end{equation}
where $rank_i$ is the index of the correctly predicted POI in the order list. This metric calculates the mean reciprocal rank of the ground-truth POIs, effectively emphasizing models that achieve higher placements in the ranked list.

%\section{Experiment Result and Analysis}
\subsection{Main Results}
\label{sec: results}
Table~\ref{tab:results} presents the performance comparison between our proposed method and various baselines across three benchmark datasets. Our method consistently outperforms all baselines, including those based on large language models (LLMs), on all datasets. In particular, compared to the strongest traditional baseline ROTAN, we achieve relative Acc@1 improvements of 23.86\%, 31.77\%, and 17.47\% on the NYC, TKY, and CA datasets, respectively. Moreover, all LLM-based methods surpass traditional approaches, indicating that the inherent commonsense knowledge in LLMs helps alleviate the sparsity issue in check-in trajectory data, making them more suitable for POI recommendation tasks. 

Additionally, compared to the strongest LLM-based baseline GNPR, our method achieves improvements of 6.33\%, 5.78\%, and 7.49\% on the NYC, TKY, and CA datasets, respectively. These improvements mainly stem from our reasoning-enhanced approach, which endows the model with a thinking process that strengthens its predictive capability. Furthermore, our method incorporates heuristic optimization techniques at both the sampling level and the token level from the perspectives of prompt and model, effectively mitigating entropy collapse and ensuring a balanced trade-off between exploration and exploitation during reasoning optimization.

\begin{table*}
\centering
\caption{Cross three datasets performance comparison. Best results are reported in bold, with the second-best results underlined.}
\begin{tabular}{lcccccccccccccc}
\toprule  
\multirow{2}{*}{\textbf{Model}} & \multicolumn{4}{c}{\textbf{NYC}} & \multicolumn{4}{c}{\textbf{TKY}} & \multicolumn{4}{c}{\textbf{CA}} \\ 
 \cmidrule(lr){2-5} \cmidrule(lr){6-9} \cmidrule(lr){10-13}
& Acc@1 & Acc@5 & Acc@10 & MRR & Acc@1 & Acc@5 & Acc@10 & MRR & Acc@1 & Acc@5 & Acc@10 & MRR\\
\midrule
PRME & 0.1159 & 0.2236 & 0.3105 & 0.1712 & 0.1052 & 0.2278 & 0.2944 & 0.1786  & 0.0521 & 0.1034 & 0.1425 & 0.1002  \\
LSTM & 0.1305 & 0.2719 & 0.3283 & 0.1857  & 0.1335 & 0.2728 & 0.3277 & 0.1834  & 0.0665 & 0.1306 & 0.1784 & 0.1201  \\
PLSPL & 0.1917 & 0.3678 & 0.4523 & 0.2806  & 0.1889 & 0.3523 & 0.4150 & 0.2542  & 0.1072 & 0.2278 & 0.2995 & 0.1847 \\
STAN & 0.2231 & 0.4582 & 0.5734 & 0.3253  & 0.1963 & 0.3798 & 0.4464 & 0.2852  & 0.1104 & 0.2348 & 0.3018 & 0.1869  \\
GETNext & 0.2406 & 0.4815 & 0.5811 & 0.3528  & 0.1829 & 0.4045 & 0.4961 & 0.2853  & 0.1526 & 0.3278 & 0.3946 & 0.2364  \\
STHGCN & 0.2734 & 0.5361 & 0.6244 & 0.3915 & 0.2950 & \underline{0.5207} & \underline{0.5980} & 0.3986  & 0.1730 & 0.3529 & 0.4191 & 0.2558  \\
TPG & 0.2555 & 0.5005 & 0.5932 & 0.3669  & 0.1420 & 0.3631 & 0.4492 & 0.2436  & 0.1749 & 0.3285 & 0.3860 & 0.2479  \\
ROTAN & 0.3106 & 0.5281 & 0.6131 & 0.4104  & 0.2458 & 0.4626 & 0.5392 & 0.3475  & 0.2199 & 0.3718 & \underline{0.4334} & \underline{0.2931}  \\ 
\midrule
LLM-Mob & 0.2582 & 0.5066 & 0.5973 & 0.3694  & 0.1673 & 0.3837 & 0.4786 & 0.2519  & 0.1764 & 0.3658 & 0.4237 & 0.2596 \\
LLM4POI & 0.3372 & - & - & -  & 0.3035 & - & - & -  & 0.2065 & - & - & - \\
Refine-POI & 0.3469 & \underline{0.5984} & \underline{0.6627} & \underline{0.4520}  & \underline{0.3160}  & 0.5140 & 0.5830 & \underline{0.4020}  & 0.2174  & \underline{0.3758} & 0.4321 & 0.2903  \\
GNPR  & \underline{0.3618} & - & - & -  & 0.3062 & - & - & - & \underline{0.2403} & - & - & - \\
\midrule
\textbf{Think2Go} & \textbf{0.3847} & \textbf{0.6287} & \textbf{0.6835} & \textbf{0.4663}  & \textbf{0.3239} & \textbf{0.5461} & \textbf{0.6051} & \textbf{0.4147}  & \textbf{0.2583} & \textbf{0.4190} & \textbf{0.4682} & \textbf{0.3167}\\
\bottomrule
\end{tabular}
\label{tab:results}
\end{table*}

\subsection{Ablation Study}
\label{sec: ablation}
To further verify the effectiveness of each key component in our approach, we conduct the following ablation studies on the three datasets, as shown in the Table:
(1) \textbf{w/o Reasoning Tokens}: We remove the reasoning tokens during fine-tuning and still employ Calibrated Advantage Policy Optimization to train the model
(2) \textbf{w/o STEU}: We exclude the spatial-temporal epistemic uncertainty weighting factor in the policy optimization loss.
(3) \textbf{w/o DRG}: We remove the difficulty-aware reward sparsity regularization term that guides the model using a full-score reward signal.
(4) \textbf{w/o Token Entropy Bonus}: We eliminate the influence of token-level entropy on the advantage estimation.

Table~\ref{tab:ablation} demonstrates that all components of our proposed method are effective across all three benchmark datasets. (1) We find that even without the use of reasoning tokens, optimizing listwise semantic IDs achieves better performance than the pointwise semantic IDs used in GNPR. This improvement is mainly attributed to the correlation among semantic IDs learned through RQ-VAE, which allows the listwise IDs to leverage the autoregressive prediction advantages of LLMs and thus reduce the difficulty of POI prediction.
(2) w/o STEU assigns higher advantage weights to harder and less predictable targets in irregular trajectory data, helping the model tackle difficult cases more effectively. (3) w/o DRG removes the reward-guided sparsity constraint, which otherwise helps guide the model using high-reward samples and alleviates entropy collapse. (4) w/o Token Entropy Bonus removes the bonus for high-entropy tokens, which otherwise promotes exploration of reasoning tokens during training.

Overall, these components contribute to the RL–based reasoning framework by enhancing the utilization and exploration of rollout data, thereby leading to improved predictive performance.

\begin{table}
\centering
\caption{Results of ablation models on NYC, TKY, and CA datasets with Acc@1 metric.}
\resizebox{1.0\columnwidth}{!}{
\begin{tabular}{lccc}
\toprule
\textbf{Ablations} &  \textbf{NYC} & \textbf{TKY} & \textbf{CA} \\
\midrule
Think2Go & 0.3847 & 0.3239 & 0.2583\\
w/o Reasoning Tokens & 0.3674 ($\downarrow 4.5\%$)& 0.3103 ($\downarrow 4.2\%$)&0.2478($\downarrow 4.07\%$)\\
w/o STEU & 0.3763 ($\downarrow 2.18\%$)& 0.3161 ($\downarrow 2.41\%$)& 0.2503($\downarrow 3.1\%$)\\
w/o DRG & 0.3744 ($\downarrow 2.68\%$)& 0.3147 ($\downarrow 2.84\%$)&0.2489 ($\downarrow 3.64\%$)\\
w/o Token Entropy Bonus & 0.3823 ($\downarrow 0.62\%$)& 0.3207 ($\downarrow 0.99\%$)& 0.2561($\downarrow 0.85\%$)\\
\bottomrule
\end{tabular}
}
\label{tab:ablation}
\end{table}

\begin{figure}
    \centering
    \includegraphics[width=1.0\linewidth]{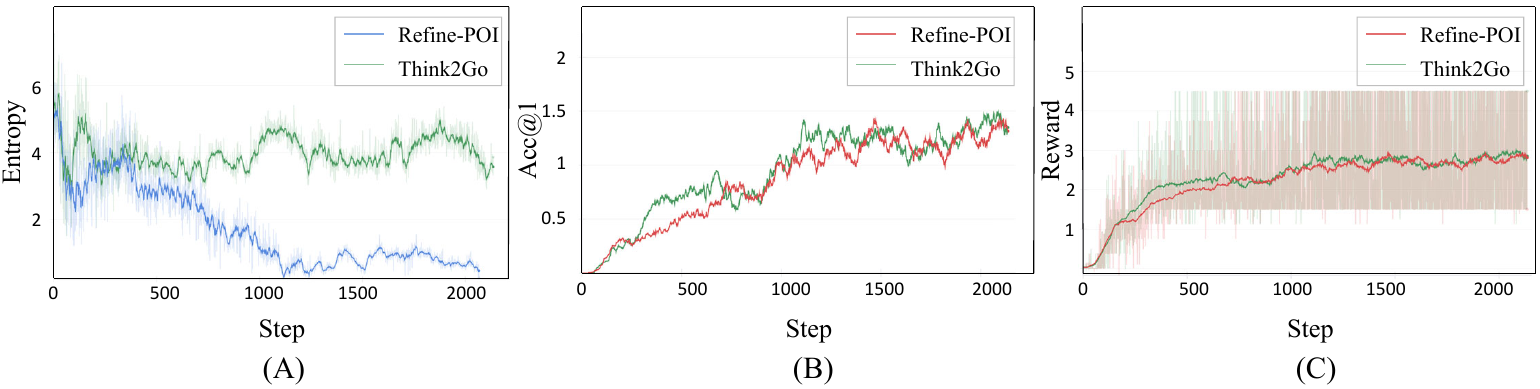}
    \caption{Advantage Estimation Analysis. We compare Refine-POI, which uses the standard GRPO method, with Think2Go, and analyze the training dynamics of Entropy, Acc@1, and Reward under each approach.}
    \label{fig:estimation}
\end{figure}

\subsection{Analysis on Advantage Estimation}
Accurate advantage estimation is critical for mitigating entropy collapse and finding globally optimal solutions in policy gradient RL. To assess this, we compare our method with Refine-POI, focusing on the behavior of their advantage estimators during training in the Figure~\ref{fig:estimation}. 

We report entropy, Acc@1, and reward over training steps, and summarize the following findings:
(1) \textbf{Entropy stability:} Think2Go better mitigates entropy collapse than Refine-POI, enabling sustained exploration. This improvement stems from the DRG, which adaptively scales the advantage based on reward signals—assigning smaller updates to incorrect predictions and maintaining higher entropy levels. We also observe that entropy first drops sharply, then rises and stabilizes. This is due to the base model initially lacking knowledge of semantic IDs and answer templates; once such knowledge is acquired, entropy increases as the model explores more diverse outputs.
(2) \textbf{Accuracy and reward performance:} Think2Go consistently outperforms Refine-POI in both Acc@1 and reward throughout training, demonstrating that its advantage estimator facilitates more effective policy optimization.
(3) \textbf{Template generalization:} The consistently high reward (above 1.5) after around 400 training steps suggests that the model has effectively learned to produce structurally correct responses. This is largely attributed to the supervision from the SFT loss, which enforces template-conforming output.

\subsection{Analysis on Reasoning-Enhanced Semantic ID}
\label{sec:reasoning-enhanced}
The reasoning-enhanced model improves prediction performance, which raises an important question: \textit{Does reasoning training enhance the model’s understanding and generalization of SID?}

To assess the LLM's semantic understanding of semantic IDs, we design a category prediction task comparing our model with the GNPR baseline trained only with SFT. Given a semantic ID, the model is asked to identify its correct category from ten randomly sampled candidates, including the ground-truth category.
As shown in the Table~\ref{tab:semantic_id}, our method achieves higher classification accuracy on SID-category prediction, confirming that reasoning training indeed improves the model’s semantic understanding and generalization of SID by expanding computation via test-time scaling. 

\begin{table}
\centering
\caption{Semantic ID prediction categories based on Think2Go (reasoning model) and GNPR (SFT model).}
\resizebox{1.0\columnwidth}{!}{
\begin{tabular}{lccc}
\toprule
\textbf{Model} &  \textbf{NYC} & \textbf{TKY} & \textbf{CA} \\
\midrule
Think2Go & 0.3024 & 0.2673 & 0.1722\\
GNPR & 0.1241 ($\downarrow 58.96\%$)& 0.1038 ($\downarrow 61.15\%$)&0.1182($\downarrow 31.36\%$)\\
\bottomrule
\end{tabular}
}
\label{tab:semantic_id}
\end{table}

\subsection{Performance on Out-of-domain Data}
We propose Think2Go, an advantage-calibrated reasoning model that improves generalization by increasing the computational depth of the reasoning process. Compared to existing methods such as GNPR, which leverage semantic IDs during fine-tuning to help the model memorize semantic information, Think2Go addresses a key limitation: in GNPR, the training of semantic IDs is decoupled from the model itself, leading to insufficient understanding of semantic IDs and poor generalization to unseen POIs.
As shown in the Table \ref{tab:generalization}, Think2Go exhibits smaller performance degradation on out-of-domain datasets compared to other methods, further demonstrating its superior generalization ability.

\begin{table}[htbp]
\centering
\caption{Cross-domain generalization comparison among \texttt{LLM4POI}, \texttt{GNPR}, and \texttt{Think2Go}. Each column shows source-to-target domain transfer performance.}
\resizebox{1.0\columnwidth}{!}{
\begin{tabular}{lcccccc}
\toprule
\textbf{Model} & \textbf{NYC-TKY} & \textbf{NYC-CA} & \textbf{TKY-NYC} & \textbf{TKY-CA} & \textbf{CA-NYC} & \textbf{CA-TKY} \\ 
\midrule
LLM4POI & 0.2594 & 0.1885 & 0.3463 & 0.1960 & 0.3344 & 0.2600 \\
GNPR    & 0.2946 & 0.2297 & 0.3481 & 0.2342 & 0.3536 & 0.2869 \\
Think2Go & \textbf{0.3183} & \textbf{0.2484} & \textbf{0.3523} & \textbf{0.2545} & \textbf{0.3677} & \textbf{0.3042} \\
\bottomrule
\end{tabular}
}
\label{tab:generalization}
\end{table}

\section{Conclusion}
In this work, we study a novel generative next POI recommendation framework with reasoning, aiming to improve model generalization by increasing the computational depth of inference. We introduce a unified fine-tuning and reinforcement learning–based reasoning model to enhance the model’s reasoning ability, where the policy is optimized by calibrating the advantage based on both the epistemic uncertainty of the data and the difficulty agnostic bias in the model's advantage estimation.
Building upon this design, we propose Think2Go, an advantage-calibrated reasoning model for time-sensitive next POI recommendation. Think2Go employs kernel density estimation to quantify epistemic uncertainty by measuring the spatial-temporal proximity between the current query and the user’s historical activity patterns. Moreover, difficulty-aware calibration enables reward-sensitive advantage estimation by scaling rewards relative to the maximum, thus mitigating bias from uniform update strength. This allows for more fine-grained and adaptive control over the policy updates, ensuring that samples of varying difficulty contribute appropriately to the learning process.
Extensive experiments and ablation studies validate the effectiveness of our proposed CAPO method, demonstrating its superior generalization capability.

\begin{acks}
This work was supported in part by the National Social Science Fund of China (No. 22BZX025) and the National Natural Science Foundation of China (No. 62572361 and 62506134). HW Qian is supported by the A*STAR Career Development Fund <Project No. C243512010>. Zhuang Zhuang is supported by the China Scholarship Council <Project No. 202406060124>.
\end{acks}

%\newpage

%\clearpage
\bibliographystyle{ACM-Reference-Format} 
\bibliography{sample-base}

%\newpage
\clearpage
\appendix

\section{Implementation Detail} 
For the fine-tuning module, we utilized LLaMA3.1-8B as the base model, fine-tuning it with LoRA (Low-Rank Adaptation) using a rank of 8, an alpha of 16, and a dropout of 0.1. The maximum generation length for tokens was set to 2048, with a batch size of 1 per GPU. The learning rate was configured to 2e-5, incorporating a 20-step warm-up phase.
For generating the reasoning module, we used VLLM for the inference module, setting its maximum generation length to 1024, a temperature of 1.3, and returning 4 rollout samples with a beam search width of 10. %During the validation phases, we also used VLLM for generating inference answers, but with more conservative parameters: a temperature of 0.7 and a beam search width of 5.
Our hardware setup comprised five Nvidia L40s GPUs. Four of these GPUs were dedicated to data training, while the remaining one was equipped with the VLLM model to generate online sampling data. Each input prompt consisted of the user's 50 most recent check-in records. During inference, we employ beam search with a size of 15 to generate multiple candidate predictions.

\section{Hyperparameter Analysis}
We conducted a parameter sensitivity study to analyze the impact of the number of sampling groups and the scaling factor $\gamma$ for optimization. As shown in the figure, the performance improves as the number of groups increases, reaching the best result when the group size $G=4$, and then stabilizing. Considering the trade-off between computational efficiency and performance, we choose $G=4$ as the default setting. In addition, $\gamma$ serves as a crucial hyperparameter that modulates the influence of spatial-temporal cognitive uncertainty on the advantage estimation. We observe that the best performance is achieved when $\gamma$ is set to 0.5.

\begin{table}[ht]
\centering
%\small  % 字体缩小
\caption{Hyperparameter analysis on group size $G$  and advantage estimator scaling coefficient $\gamma$.}
\begin{tabular}{cccc}
\toprule
\textbf{Hyperparameter} &  \textbf{NYC} & \textbf{TKY} & \textbf{CA} \\
\midrule
$G$=2 & 0.3785 & 0.3217 & 0.2538\\
$G$=3 & 0.3802 & 0.3224 &0.2551\\
$G$=4 & 0.3847 & 0.3239 & 0.2583\\
$G$=5 & 0.3840 & 0.3246 & 0.2580\\
\midrule
$\gamma=0.3$ & 0.3803 & 0.3209 & 0.2524\\
$\gamma=0.4$ & 0.3819 & 0.3221 &0.2567 \\
$\gamma=0.5$ & 0.3847 & 0.3239 & 0.2583\\
$\gamma=0.6$ & 0.3810 & 0.3217 & 0.2554\\
$\gamma=0.7$ & 0.3742 & 0.3198 & 0.2511\\
\bottomrule
\end{tabular}
\label{tab:parameter}
\end{table}

\section{Prompt Format}

\subsection{Reasoning Prompt Format}
\label{A.2.1}
We provide our reasoning prompt template in the Table~\ref{tab:reasoning_prompt}. It is important to note that the training and inference stages must remain consistent to ensure the model produces correct reasoning and prediction formats.

\begin{table}[ht]
  \centering
  \caption{The prompt format of reasoning for training and inference.}
  \renewcommand{\arraystretch}{1.1}
  \setlength{\tabcolsep}{3pt}
  \begin{tabularx}{\columnwidth}{>{\bfseries}m{1.8cm} X}
    \hline
    System \newline Instruction &
    You are a helpful assistant specialized in location prediction. You analyze POI access patterns to predict next locations. Respond in the following format: \texttt{<think>\textbackslash n[Your step-by-step reasoning here]\textbackslash n</think>\textbackslash n<answer>: <a\_7><b\_25><c\_16></answer><|end\_of\_text|>} \\
    \hline
    Instruction &
    Here is a record of a user's POI check-in trajectory. Your task is to analyze the user's POI check-in trajectory patterns and predict the next location they will visit at the specified time. You should perform a detailed analysis of temporal patterns, transitions between POIs, and user preferences. Present your reasoning step-by-step inside a \texttt{<think>} and \texttt{</think>} block. After completing your analysis, provide your final POI prediction within \texttt{<answer>:} and \texttt{</answer>} tags. For example: \texttt{<answer>: <a\_7><b\_25><c\_16></answer><|end\_of\_text|>} \\
    \hline
    Self-correct &
    Unfortunately (Fortunately), the predicted answer is incorrect (correct). It is more likely that the user will visit: \\
    \hline
    Input &
    User\_\textcolor{green}{<uid>} visited: \textcolor{blue}{<SID>} at \textcolor{orange}{[time]}, ..., \textcolor{blue}{<SID>} at \textcolor{orange}{[time]}. When \textcolor{red}{[time]} user\_\textcolor{green}{<uid>} is likely to visit: \\
    \hline
    Example &
    User\_3 visited: \texttt{<a\_22><b\_20><c\_25><d\_0>} at 2012-05-05 13:58, \texttt{<a\_22><b\_18><c\_16><d\_0>} ..., \texttt{<a\_6><b\_27><c\_31>} at 2012-12-10 10:07. When 2012-12-30 20:47 user\_3 is likely to visit: \\
    \hline
  \end{tabularx}
  \label{tab:reasoning_prompt}
\end{table}

\subsection{Zero-shot Semantic Information Prediction Prompt Format for SID}
\label{A.2.2}
In the Table~\ref{tab:reasoning_enhanced}, we demonstrate that semantic IDs possess semantic information after fine-tuning with SFT and Think2Go reinforcement reasoning. To help the model predict categories in the correct format under zero-shot conditions, we need to customize the prompt template.% and provide an example for illustration. 
Specifically, to guide the model in predicting the category corresponding to the target semantic ID, we employ a concrete example as a prompt.

\section{Case Study}
To more intuitively demonstrate Think2Go's reasoning behavior, we analyzed representative cases from the NYC dataset in Table~\ref{tab:case}, including a real example from user ID 451 in the test set. It shows how LLM utilizes common knowledge combined with prompts to analyze the target check-in process based on the semantic ID transition patterns and temporal periodicity of sequential check-ins. This helps the LLM understand semantic IDs, rather than just simple symbolic transition patterns.

\begin{table}[H]
  \centering
  \caption{The prompt format for zero-shot semantic information prediction.}
  \renewcommand{\arraystretch}{1.1}
  \setlength{\tabcolsep}{3pt}       % 减小列间距
  \begin{tabularx}{\columnwidth}{>{\bfseries}m{1.8cm} X}
    \hline
    System \newline Instruction & You are a location prediction assistant. Predict POI categories based on user check-in patterns. Output only category names, not semantic IDs.  \\
    \hline
    Instruction & Here is a reference example to guide your response: [example]
    Trajectory: [Input] 
    Now, you are required to select the category corresponding to the semantic ID of the target POI from the following list of candidate categories: 1.epartment Store, 2.Comedy Club, 3.Racetrack, 4.City, 5.Cosmetics Shop, 6. Building, 7.Plaza, 8.Bookstore, 9.Nursery School, 10.Light Rail.
    <answer>: \\
    \hline
    Input & User\_\textcolor{green}{<uid>} visited: \textcolor{blue}{<SID>} at \textcolor{orange}{[time]}, ..., \textcolor{blue}{<SID>} at \textcolor{orange}{[time]}. When \textcolor{red}{[time]} user\_\textcolor{green}{<uid>} is likely to visit: \textcolor{blue}{<SID>}. \\
    \hline
    example & Trajectory: <a\_1><b\_1><c\_29> at 2013-01-01 10:00, <a\_2><b\_3><c\_8> at 2013-01-01 12:00. 
    When 2013-01-02 20:47 user is likely to visit: <a\_2><b\_1><c\_29>.
    Now, you are required to select the category corresponding to the semantic ID of the target POI from the following list of candidate categories: 1.Food, 2.Shopping, 3.Hotel, 4.Education, 5.Park, 6.Hospital, 7.Transport, 8.Office, 9.Entertainment, 10.Residence.
    <answer>: Hospital</answer> \\
    \hline
  \end{tabularx}
  \label{tab:reasoning_enhanced}
\end{table}

\begin{table}[H]
  \centering
  \caption{The case of test examples.}
  \renewcommand{\arraystretch}{1.1}
  \setlength{\tabcolsep}{3pt}       % 减小列间距
  \begin{tabularx}{\columnwidth}{>{\bfseries}m{1.8cm} X}
    \hline
    Input & User\_451 visited:<a\_28><b\_27><c\_22><d\_2> at 2012-11-15 20:45, ..., <a\_1><b\_23><c\_20> at 2013-01-08 10:21. When 2013-01-09 07:13 user\_451 is likely to visit: \\
    \hline
    Reasoning & <think>The best predictions often emerge when one follows recurring behavioral patterns rather than attempting to account for every possible deviation. In this user’s case, visits are centered around a few high-frequency POIs, with notable repetitions during weekday mornings and afternoons. Particularly, the user tends to return to familiar places shortly after recent visits, especially those with high historical engagement. When transitions repeat with short intervals, it often signals a strong preference or necessity. In this context, recent consecutive visits to the same POI indicate a reinforced habit. Visits in the early morning typically follow the previous day's routine, often resuming at the same or nearby POI. Considering all these patterns, it is highly probable that the user will repeat the last known check-in behavior.</think>
    <answer>: <a\_1><b\_23><c\_20></answer><|end\_of\_text|> \\
    \hline
    Answer & <a\_1><b\_23><c\_20> \\
    \hline
  \end{tabularx}
  \label{tab:case}
\end{table}

\section{Analysis on Model Efficiency}
In this paper, we propose Think2Go, which unifies SFT and RL-based reasoning models. We utilize semantic IDs as data and only incorporate temporal information to describe check-in trajectories. 
Compared to conventional RL-based reasoning methods, this approach achieves higher efficiency. Table~\ref{tab:efficiency} reports a comparison of the efficiency between our method and Refine-POI (RL methods based on standard text prompts). Both methods were trained on the NYC dataset. It can be observed that our method achieves higher efficiency in both training and testing time and reduces memory consumption. This improvement benefits from using semantic IDs as POI identifiers, which reduces redundant input information, and from online sampling with VLLM during the reasoning stage, which also helps accelerate sampling speed.

\begin{table}[H]
\centering
\caption{Efficiency comparison on the NYC dataset.}
\resizebox{1.0\columnwidth}{!}{
\begin{tabular}{cccc}
    \toprule
    \textbf{Models}&     \textbf{Training Time per epoch}& \textbf{Test Time}& \textbf{Memory}\\
    \midrule
    Refine-POI &  6.2h/epoch & 1.4h & 632GB\\
    Think2Go & 2.7h/epoch ($\downarrow56.45\%$)& 0.5h ($\downarrow64.29\%$)& 228GB($\downarrow63.92\%$)\\
    \bottomrule
\end{tabular}
}
\label{tab:efficiency}
\end{table}

\end{document}